\documentclass[12pt,journal,compsoc]{IEEEtran}

\ifCLASSINFOpdf
\else
\fi
\usepackage[dvips]{graphicx}

\ifCLASSOPTIONcompsoc
   \usepackage[nocompress]{cite}
\else
   \usepackage{soul}
\fi

\ifCLASSINFOpdf
\else
\fi

\begin{document}
%
\title{Thermal Control of a Dual Mode\\  
Parametric Sapphire Transducer}

\author{Jacopo~Belfi,
			Nicol\`o~Beverini,
		  Andrea~De~Michele,
		  Gianluca~Gabbriellini,
      Francesco~Mango and Roberto~Passaquieti
\IEEEcompsocitemizethanks{\IEEEcompsocthanksitem The authors are with the Department of Physics ``Enrico Fermi'' and \textit{CNISM} unit\`a di Pisa , Universit\`a di Pisa,
56127 Pisa, Italy.\protect\\
E-mail: belfi@df.unipi.it}}
 
\ifCLASSOPTIONpeerreview
\markboth{Journal of \LaTeX\ Class Files,~Vol.~6, No.~1, January~2007}%
{Shell \MakeLowercase{\textit{et al.}}: Bare Advanced Demo of IEEEtran.cls for Journals}
\fi
\IEEEcompsoctitleabstractindextext{
\begin{abstract}
We propose a method to  control the thermal stability of a sapphire dielectric 
transducer made with two dielectric disks separated by a thin gap and resonating 
in the whispering gallery (WG) modes of the electromagnetic field. 
The simultaneous measurement of the frequencies  of both a WGH mode and a WGE mode 
allows one to discriminate the frequency shifts due to gap variations  from those due to temperature instability.  
A simple model, valid in quasi equilibrium conditions, describes the frequency shift of the two modes in terms of four 
tuning parameters. A procedure for the direct measurement of them is presented. 
\end{abstract}}
\maketitle

\IEEEdisplaynotcompsoctitleabstractindextext

 \ifCLASSOPTIONpeerreview
 \begin{center} \bfseries EDICS Category: 3-BBND \end{center}
 \fi
%
\IEEEpeerreviewmaketitle

\section{Introduction}
\IEEEPARstart{A}{dielectric} whispering gallery resonator, made with two dielectric disks separated by a thin
gap is a very sensitive displacement sensor \cite{Peng_92}. 
Indeed, parametric microwave sapphire oscillators, operating at cryogenic temperature, have been proposed as readout systems for bar-type gravitational antennae \cite{IEEE45_5_1998} and also as core sensors for space-gravity tests and geodesy research \cite{J_phys_D_27_1994}.
It has been shown that such devices provide  high sensitive vibration measurements even when operating at room-temperature \cite{Peng_94}.

The efficiency of these transducers
is conveniently described by the merit factor $$M=\frac{Q\cdot\partial_z f}{f},$$ where $Q$ is the resonator
quality factor, $f$ its resonance frequency and $\partial_z f$ is the tuning coefficient, the derivative of the resonance frequency w.r.t. the gap spacing $z$. 

The modes are classified as $WGH$ modes (characterized by $E_z,
H_{\phi}, H_{\rho})$ and $WGE$ modes (with $H_z,
E_{\phi}, E_{\rho})$, where $z$ denotes the  component of the field vector along the disks rotational axis, $\rho$ the radial component, and $\phi$ the azimuthal one.

The $WGH$ modes with high azimuthal mode number  present the highest merit factors \cite{IEEE45_5_1998}.

Even at room temperature, X-band  resonances in a sapphire resonator made with  disks 4 cm in diameter and 0.5 cm in thickness, exhibit a $Q$ factor of $\sim10^5$ and a tuning coefficient of $6\,\, \rm{MHz}~\rm{\mu m}^{-1}$ so that  $M\sim60~\rm{\mu m}^{-1}$. The frequency instability is dominated by thermal effects, which contribute with some tenths of $\rm{MHz}~\rm{K}^{-1}$.
This strong temperature dependence of the dielectric tensor  compromises the long term stability. On the other side, ultra-low frequency (i.e. diurnal timescale) displacement measurements have a key role in many applications such as gravimetric exploration, environmental monitoring and materials testing.\\
In the research field of ultra stable oscillators, several techniques for precision temperature stabilization have been proposed.
At low temperature (below 100 K) the cancellation (to first order) of the temperature coefficient of frequency can be achieved by employing doped dielectric disks \cite{doped}, composite sapphire-rutile resonators \cite{composite} and mechanically compensated structures \cite{mec_comp}. 
High  thermal stability at room temperature can be obtained with high precision control stages, by the optimization of temperature sensors \cite{sensor} and actuators \cite{actuator}, and also by accurately modelling and simulating the thermodynamic system under test \cite{termico_boudot}.

Due to the anisotropy of the sapphire crystal, the temperature coefficient of frequency is in general
different for WGH and WGE modes. Exciting two  electromagnetic modes with different polarization in the same
resonator permit one to measure and stabilize the resonator temperature in oscillators \cite{Dual1,Dual2} even at room temperature \cite{Dual3}.
In the case of a displacement transducer, WGH and WGE modes have to satisfy different boundary conditions at the gap spacing thus  exhibiting  also  different tuning coefficients $\partial_z f$. 

The measured resonance frequency variations in a parametric sapphire transducer are given by the mixing of  \textit{pure-displacement} signals, \textit{pure-temperature} signals and \textit{temperature-induced displacement} signals. These last are due to  the  thermal expansion of the material comprising the  enclosing chamber.\\
In this paper we propose a calibration technique providing an estimate of pure displacement signals in a dual mode parametric sapphire resonator transducer operating at room temperature.
\begin{figure}[!h]
\centering
\includegraphics[width=3.4in]{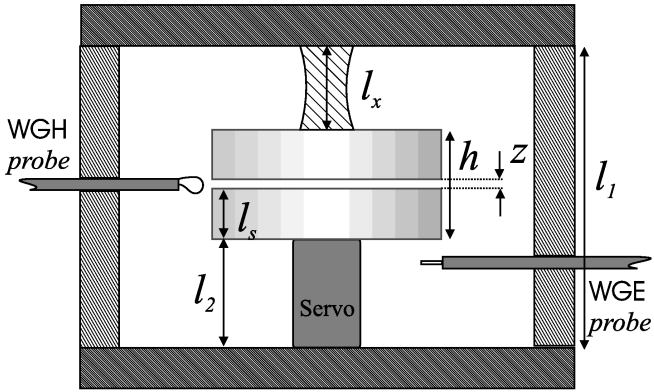}
\caption{A generic displacement sensor based on a transducer converting the physical quantity ``x'', in a vertical displacement $\delta l_x$. A servo actuator controls the separation set point. The sapphire disks are cut with their c-axis parallel to the cylinder axis.}\label{section}
\end{figure}
\section{Thermal effects in a Sapphire Parametric Transducer}
The frequency dependence on temperature for the parametric displacement transducer of Fig.~\ref{section} is described by the following equation~\cite{Dual2}:
\begin{eqnarray}
&&\frac{1}{f}\frac{\partial f}{\partial T}=-p_{\epsilon\bot}\alpha _{\epsilon\bot}
-p_{\epsilon||}\alpha _{\epsilon||}\\\nonumber&&-p_D \alpha_D-p_{h} \alpha_{h}-p_z (2\alpha_{l_s}+\alpha_{l_2}+\alpha_{l_x}-\alpha_{l_1}),
\end{eqnarray}
where $\alpha _{\theta}=\frac{1}{\theta}\frac{\partial \theta}{\partial T}$ are coefficients depending on the materials  and $p _{\theta}=\frac{\theta}{f}\frac{\partial f}{\partial \theta}$ are the filling factors of the considered WG mode and depend on the field distribution of the mode inside the resonating volume. The  label $\theta$ refers respectively to: $\epsilon_{\bot}$ and $\epsilon_{||}$ the dielectric constants perpendicular and parallel to the c-axis, $D$ and $h$ the resonator spatial dimensions  perpendicular and parallel to the c-axis, $l_s$, $l_1$, $l_2$, $l_x$, $z$ the vertical dimensions of the single sapphire disk, the chamber length, the lower support (actuator), the upper support (moving part) and the gap spacing. 
Under stationary conditions, one can assume to be valid the following linear and time-independent relation between a given pair of WG modes, the temperature $T$ and the gap spacing $z$: 

\begin{equation}
\left(\begin{array}{c}
\delta f^{WGH}\\\delta f^{WGE}
\end{array}\right)=
\textbf{C}\left(\begin{array}{c}
\delta T\\\delta z
\end{array}\right)
\end{equation}
with
\begin{equation}\textbf{C}=
\left(\begin{array}{c}
C_T^{WGH} \,\,\,\, C_z^{WGH}\\
C_T^{WGE} \,\,\,\,  C_z^{WGE}
\end{array}\right).
\end{equation}
The anisotropy of both the material and the field distribution assures that it is possible to invert  $\textbf{C}$ and  to obtain the following estimate for the effective temperature fluctuations $\delta T^*$ and for the pure (not thermally-induced) displacement $\delta z^*$:

\begin{equation}
\left(\begin{array}{c}
\delta T^*\\\delta z^*
\end{array}\right)=
\textbf{C}^{-1}
\left(\begin{array}{c}
\delta f^{WGH}\\\delta f^{WGE}
\end{array}\right).
\end{equation}

\section{Calibration}
\begin{figure}
\begin{center}
  \includegraphics[width=3.4in]{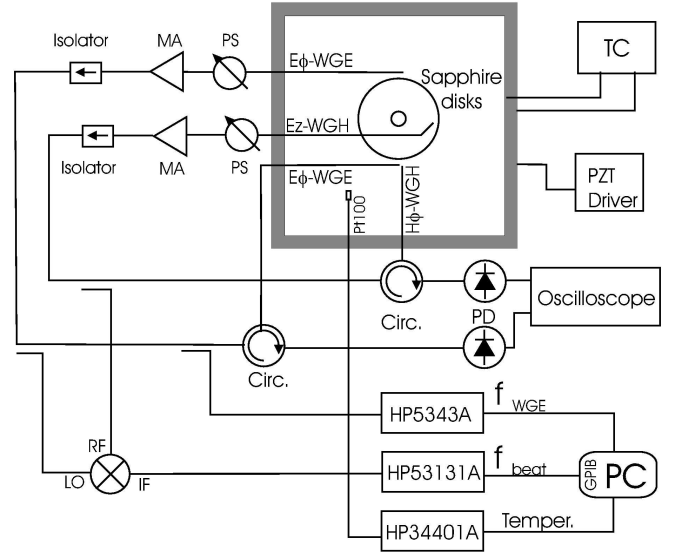}
  \end{center}
    \caption{Microwave circuitry for the dual mode self oscillating transducer employed for the calibration. TC: Temperature Controller, PS: Phase Shifter, MA: Microwave Amplifier (ALC-ALN060029), Circ: Circulator, PD: Power Detector.}\label{setup_TETM}
    \end{figure}

\begin{figure}
\begin{center}
  \includegraphics[width=3.3in]{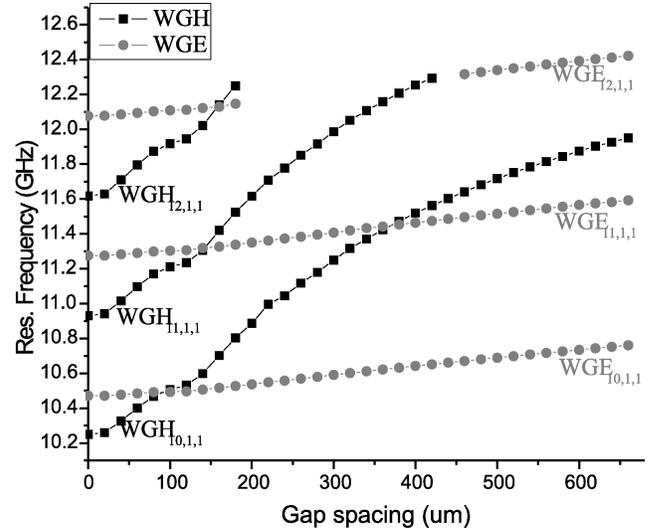}
  \end{center}
    \caption{Resonance frequencies of WGH and WGE modes versus gap spacing.}\label{TE_TM_z}
\end{figure}

The basis of the  method is to determine the coefficients of the matrix $\textbf{C}$ by means of a calibration procedure.
A schematics of the  experimental apparatus for the sensor calibration is shown in Fig.~\ref{setup_TETM}.
The sapphire transducer is placed inside a metallic chamber which is temperature stabilized at about $34^\circ\rm{C}$. Each chamber wall is electrically isolated  in order to avoid any influence from cavity modes on the chosen  WG modes in the dielectric.
A 4 cm thick insulation material covers the whole chamber to reduce the heat losses.
The temperature stabilization system consists of a standard PI controller driving a set of thermoresistances in thermal contact with the metallic cavity.
The set point temperature is determined by a voltage input for calibration purposes. 
The sapphire disks composing the resonator transducer are coupled to four microwaves antennae (see Fig.~\ref{setup_TETM}). Two electric probes, coupled to the azimuthal electric field are used for sustaining the WGE oscillations. One electric probe, coupled to the axial electric field, and one magnetic probe, coupled to the azimuthal magnetic field, excite the WGH oscillations.
We chose to test the system using the following modes (see Fig.~\ref{TE_TM_z}):
\begin{eqnarray}
f_{WGE_{11,1,1}}&\sim&11.38\,\, \rm{GHz}\,\,,\\\nonumber
f_{WGH_{10,1,1}}&\sim&11.20\,\, \rm{GHz}\,\,,
\end{eqnarray}
for $z\sim 300\,\, \rm{\mu m}$.
In this choice there is very little coupling \cite{Peng_92} between the selected modes and furthermore they are close enough in frequency so that it is possible to measure their beat note by means of an RF frequency counter.
The upper resonator  disk is rigidly attached to the aluminium top plate and the lower disk is mounted on a piezoelectric actuator.

A PC based data acquisition system monitors three physical quantities:
internal temperature (via the resistance of a Pt100 temperature probe), the $WGE$ mode frequency (with a microwave frequency counter) and the $WGE-WGH$ beat frequency (with an RF frequency counter).

\section{Check of the method}
Once the two microwave oscillations are implemented, we can  evaluate the four elements of the $\textbf{C}$ matrix by means of the temperature and position controllers.
$C_z^{WGE}$ and  $C_z^{WGH}$ are given by the ratio between the frequency variation of $f^{WGE}$ and $f^{WGH}$ and the calibrated gap variation induced by the piezoelectric  actuator moving the lower disk.

\begin{figure}
\begin{center}
\includegraphics[width=3.3in]{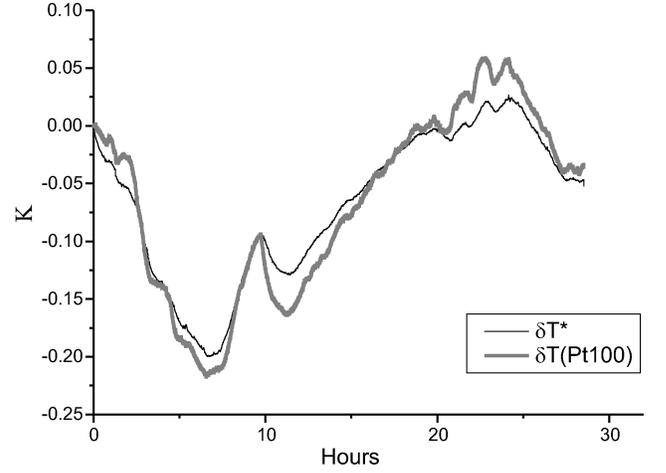}
\end{center}
\caption{Comparison between frequency-based temperature estimate and temperature measured by a Pt100 probe.}\label{cnfr_temp}
\end{figure}

\begin{figure}
\begin{center}
\includegraphics[width=3.3in]{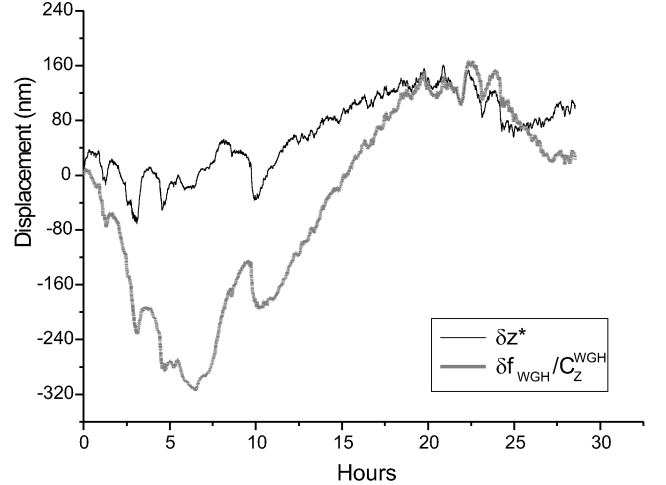}
\end{center}
\caption{Comparison between the dual frequency based displacement estimate and the estimate obtained from the single WGH mode.}\label{cnfr_z}
\end{figure}

$C_T^{WGE}$ and  $C_T^{WGH}$  are instead the proportionality constants between the frequency variation extrapolated at thermal equilibrium and the temperature variation induced by thermal actuators.
For the present setup we obtained:
\begin{eqnarray*}
&C_T^{WGE}=0.155\,\rm{MHz}/\rm{K},\\\nonumber
&C_T^{WGH}=3.64\,\rm{MHz}/{K},\\\nonumber
&C_z^{WGE}=0.567\,\rm{MHz}/\rm{\mu m},\\\nonumber
&C_z^{WGH}=3.43\,\rm{MHz}/\rm{\mu m}.
\end{eqnarray*}
In Fig.~\ref{cnfr_temp} we show  a comparison between the frequency-based temperature variation estimate $\delta T^{*}$ and the temperature variation measured by a Pt100 thermometer placed inside the chamber. The measurement is referred to stabilized temperature conditions and rigid materials.  The agreement between the two traces confirms the validity of the model especially for very slow variations and makes it possible to measure the sensor temperature sensitivity by means of frequency measurements.
In Fig.~\ref{cnfr_z} we show a comparison between the displacement estimate $\delta z^{*}$, obtained from the calibrated dual frequency measurement, and $\delta f^{WGH}/C_z^{WGH}$ i.e. the displacement estimate one would get from the $WGH$ mode alone (the most sensitive mode to the gap spacing variation).
It is neatly visible that the trace $\delta z^{*}$ displays a net reduction of the fluctuations over time scales typical of thermal phenomena.

\begin{figure}
\begin{center}
  \includegraphics[width=3.3in]{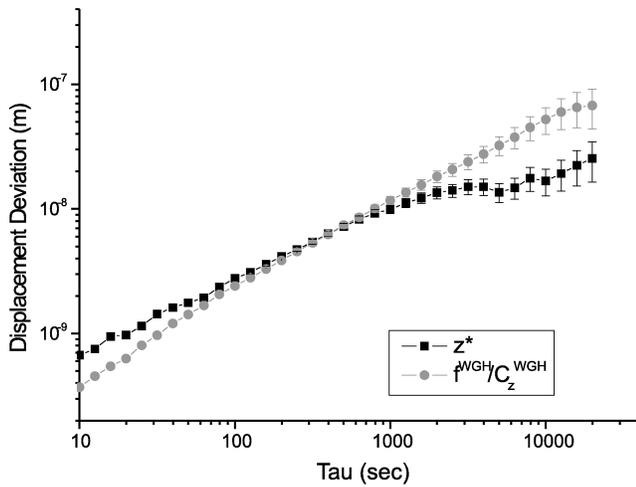}
  \end{center}
    \caption{Allan deviation of the displacement fluctuations shown in  Fig.~\ref{cnfr_z}.}\label{cnfr_allan}
\end{figure}

It is worth  remarking that the assumption of thermal equilibrium for the system is  essential to effectively filter out the temperature noise from the displacement signal. 
In Fig.~\ref{cnfr_allan}  we show the  Allan deviation  $s_z(\tau)= \sqrt{\langle(\bar{z_{i}}-\bar{z}_{i+1})^2\rangle/2}$ (where $\bar{z_{i}}$ is the average of the displacement $z$ over the i--th sample--period of duration  $\tau$) of the two displacement traces of  Fig.~\ref{cnfr_z}.

It can be seen that for integration times below about 200 sec the dual mode based displacement measurements are noisier than the single frequency measurements. This is due to the fact that over these time scales the different sensor components are not in thermal equilibrium  and the two frequencies are almost uncorrelated. For integration  times above about 500 sec, a net reduction of noise can be observed for the trace due to $\delta z^{*}$. This time scale corresponds to the longest time constant (sapphire thermalization) in the sensor. Here thermal equilibrium approximation is almost valid and then the method provides the expected noise cancellation. 

\section{Conclusions}
We proposed and tested an experimental technique for reducing the instabilities induced by thermal fluctuations in a sapphire parametric  displacement transducer. The effective temperature of the sensor and the pure displacement signal can be obtained from a double frequency measurement.
The very simple time-independent model of the system has been tested  on a trial setup in order to outline the potential and the limitations of the method.
In the future we intend to improve the efficiency of the thermal stabilization and isolation of the system in order to improve the bandwidth of the noise filtering. Finally by implementing a double  locking  of the two reconstructed quantities ($\delta T^{*}$ and $\delta z^{*}$) it will be possible to implement the experiment under static conditions and improve on the modelled assumptions.




%


\end{document}